# Operando study of the evolution of peritectic structures in metal solidification by quasi-simultaneous synchrotron X-ray diffraction and tomography

Kang Xiang [1], Yueyuan Wang [1], Shi Huang [1,2], Hongyuan Song [1], Alberto Leonardi [3], Peter Garland [3], Sharif Ahmed [3], Michał M. Kłosowski [4], Hongmei Yang [5], Mengnie Li [5,*], Jiawei Mi [1,*]

[1] *School of Engineering and Technology, University of Hull, Hull, HU6 7RX, UK*

[2] *School of Engineering and Westlake Institute for Advanced Study, Westlake University, Hangzhou, 310030, P.R. China*

[3] *Diamond Light Source Ltd., Didcot, Oxfordshire, OX11 0DE, UK*

[4] *Research Complex at Harwell, Didcot, Oxfordshire, OX11 0FA, UK*

[5] *Faculty of Material Science and Engineering, Yunnan Key Laboratory of Integrated Computational Materials Engineering for Advanced Light Alloys, Kunming University of Science and Technology, Kunming, 650093, P.R. China*

*\* Corresponding authors: J.Mi@hull.ac.uk (J. Mi), limengnie@kust.edu.cn (M. Li)*

**Abstract**

Using quasi-simultaneous synchrotron X-ray diffraction and tomography techniques, we have studied *in-situ* and in real-time the nucleation and co-growth dynamics of the peritectic structures in an Al-Mn alloy during solidification. We collected ~30 TB 4D datasets which allow us to elucidate the phases' co-growth dynamics and their spatial, crystallographic and compositional relationship. The primary $Al_4Mn$ hexagonal prisms nucleate and grow with high kinetic anisotropy -70 times faster in the axial direction than the radial direction. In all cases, a ~5 μm Mn-rich diffusion layer forms at the liquid-solid interface, creating a sharp local solute gradient that governs subsequent phase transformation. The peritectic $Al_6Mn$ phases nucleate epitaxially within this diffusion zone, initially forming a thin shell surrounding the $Al_4Mn$ with an orientation relationship of $\{10\bar{1}0\}_{HCP}$ // $\{110\}_O$, $[0001]_{HCP}$ // $[001]_O$. Such ~5 μm Mn-rich diffusion layers also cause solute depletion at the liquid side of the liquid-solid interface, limiting further epitaxial phase growth, but prompting phase re-nucleation and branching at crystal edges, resulting tetragonal prism structures that no longer follow the initial orientation relationship. The anisotropic diffusion also led to the formation of core defects at the centre of both phases. Furthermore, increasing cooling rate from 0.17 to 20 °C/s can disrupt the stability of the solute diffusion zone, effectively suppressing the formation of the core defects and forcing a transition from faceted to non-faceted morphologies. Our work establishes a new theoretical framework for how to tailor and control the peritectic structures in metallic alloys through solidification processes.

## 1. Introduction

Peritectic reaction in an alloy system is a very common phase reaction (or transformation) in the liquid-to-solid solidification processes [1-3], in which the primary solid phase (i.e., the 1$^{st}$ phase solidified from the liquid) reacts with the remaining liquid to form the 2$^{nd}$ solid phase. The resulting peritectic phase (normally an intermetallic compound) often has a complex composite-type structure coupled with convoluted 3D morphology [3-6]. The dynamic evolution of such structure and morphology is mainly determined or driven by (1) the inherent materials properties (the primary phase's structure, size and morphology, atom diffusion and liquid-solid interfacial properties); (2) the thermodynamic driving forces and/or external fields that are present at the reaction (e.g., thermal gradient, cooling rate, natural/forced convection, and magnetic fields, etc) [3, 7-9]; and (3) any other phase reactions that are directly interacting with the peritectic reaction. Normally, the crystalline structure of the primary phase has a profound effect on the nucleation and growth dynamics of the 2$^{nd}$ phase. For example, Lü *et al* [6] studied the Ni-Zr system and identified the orientation relationships between the primary Ni$_7$Zr$_2$ and peritectic Ni$_5$Zr phases as {111} Ni$_7$Zr$_2$ // {111} Ni$_5$Zr. Liu *et al [9]* investigated the Sn-Co system and found that the primary CoSn and peritectic CoSn$_2$ phases followed orientation relationships of {0001} CoSn // {110} CoSn$_2$ and {10$\bar{1}$0} CoSn // {110} CoSn$_2$. Also, refining the size of the primary phase has a dominant effect on enhancing peritectic transformation in peritectic steel systems and superconductor materials [7, 10]. Clearly, an in-depth understanding of the compositional, spatial and crystallographic orientation relationships of the peritectic phases is crucial for tailoring and controlling the final peritectic microstructures. Unfortunately, there have been very limited real-time and operando studies in this direction, especially in 4D space (3D + time).

Here, we are particularly interested in the peritectic systems with the primary phase having a hexagonal close-packed (HCP) structure and the 2$^{nd}$ phase having an orthorhombic structure. There are a large number of scientifically and technologically important alloy systems with their peritectic structures formed via such HCP to orthorhombic structure transition. For example the Al-Mn system (HCP Al$_4$Mn to orthorhombic Al$_6$Mn) [11]; the Ti-Al binary system (HCP α-Ti to orthorhombic TiAl) [12]. The thermoelectrical Zn-Sb binary system (HCP α-Zn to orthorhombic ZnSb) [13]. To the best of our knowledge, so far, there are no systematic *in-situ* and operando studies dedicated to the real-time peritectic reaction dynamics with the HCP to orthorhombic structure transition.

Here, we used an Al-Mn binary system as the model alloy and did a systematic and comprehensive operando studies on how the primary phase (Al$_4$Mn) to initiate the 2$^{nd}$ phase (Al$_6$Mn) at the peritectic reaction and the subsequent growth dynamics. We choose such system is because the Al-Mn alloy system has a series of typical peritectic reactions at certain compositions and temperature in solidification. There are multiple Al-Mn intermetallic phases ($Al_6Mn$ – HCP structure) and $Al_4Mn$ (orthorhombic structure), which are nucleated either concurrently or sequentially (especially when the Mn is > 1.6 wt.%) [14, 15]. The nucleation and co-growth of these intermetallics is driven by the intricate thermodynamic and kinetic interactions, including competitive nucleation, solute redistribution, phase impingement, etc [16, 17]. These dynamic interactions often produce complicated 3D morphologies that are very difficult, if not impossible, to be characterized or quantified by ex-situ approaches [18]. More importantly, Mn is a very unique element that is able to change or alter a series of intermetallic phases in the Al-based alloys containing Si, Mg, Cu, Zr, and Fe, etc. Particularly useful or effective in changing or altering the Fe-containing intermetallic phases [19, 20]. Thus, many attempts have been made in the past to optimise their size, morphology and distribution characteristics [20-23]. But majority of these studies have used traditional 2D metallographic techniques in characterisation, which are not able to reveal the true 3D complexity and dynamic evolution of these intermetallic phases at the peritectic reactions during solidification. Recent advances of the X-ray and electron based *in-situ* characterization techniques make it possible to obtain complementary real space and reciprocal space data for metal alloys in solidification under operando conditions [5, 24-27]. For example, simultaneous X-ray diffraction and tomography techniques can obtain the crystallographic evolution and 3D morphology of solidifying phases in real time [24-26, 28]. It is essential for studying the nucleation and growth dynamics of intermetallic compounds, tracking precisely the phases' growth and morphological evolution [25, 29]. The obtained real-time data are critical for us to understand and quantify the nucleation, growth, and transformation dynamics of different phases, which control the microstructural development during the solidification processes [25, 29-32].

In this paper, we present our recent work of using the quasi-simultaneous synchrotron X-ray diffraction and tomography techniques available at the DIAD beamline (K11) of the Diamond Light Source to study *in-situ* and in real-time the nucleation and co-growth dynamics of coupled peritectic phases and their compositional, spatial and orientation relationships in an Al-8wt.%Mn alloy during the solidification process. This systematic research work has produced ~30 TB 4D (3D space + time) datasets that allow us to fully understand the underlying mechanisms of the nucleation and co-growth dynamics of the coupled peritectic phases and the complete evolution of the peritectic reactions. These findings provide a quantitative and theoretical foundation for tailoring the architecture, and hence the properties, of peritectic alloys through precise control of solidification kinetics.

## 2. Experiments and methods
## 2.1. Alloys sample preparation

The Al-8wt.%Mn ingot was made by melting 230 g pure Al (99.96 wt.%) and 20 g pure Mn (99.85 wt.%, purchased from the Goodfellow, UK) in an alumina boat crucible (the crucible inner surface was coated with a boron nitride spray) using a tube furnace operated at 1200 °C for ~6 hours. During the heating and melting process, high-purity argon gas (flow rate of 0.6 L/min) was flowed into the tube furnace to prevent oxidation. After completely melted the charge, the alloy was cooled down with the furnace (switched off the heating power) to room temperature. Then, the ingot was placed into an electrical resistance furnace and remelted at ~800 °C. During the remelting operation, stirring of the melt was carried out in every 10 minutes to homogenize the melt temperature and composition. Finally, the melt was vacuum-sucked into a quartz tube (Φ 2 mm × 200 mm, 1 mm wall thickness) using a dedicated counter-gravity casting apparatus [33] at a negative pressure of 70 KPa, solidified inside the quartz tube to form cylindrical bar samples.

## 2.2. Quasi-simultaneous synchrotron X-ray diffraction and tomography experiments

Figs. 1a and 1b show the experimental set-up for the quasi-simultaneous synchrotron X-ray diffraction and tomography experiments at the DIAD beamline (K11) of the Diamond Light Source (DLS) [34]. For a detailed description of the quartz tube furnace, please refer to [35]. The alloy sample (Φ 1.8 × 15 mm) was put inside a capillary quartz tube (Φ 2.5, 0.1 mm wall thickness), which was set on top of an alumina tube holder (Φ 5 × 100 mm, 1.0 mm wall thickness). The capillary tube with the sample was then placed inside the quartz tube furnace chamber surrounded by an electrical-resistance heating coil (see the inset in Fig. 1a). The furnace temperature was controlled by an Omega PID thermal controller (CN16DPT-440). Three Φ 0.25 mm K-type thermocouples (labelled as TC1, TC2 and TC3 in Fig. 1a) were used to control and monitor the temperatures inside the capillary tube and the quartz furnace chamber. Because of rotating operation during tomography scans, the thermocouple was placed approximately ~3 mm above the top of the sample. A PicoTC08 data logger and PicoLog6 software were used to record the temperature. Before any *in-situ* experiment, melting of the alloy sample and control of the temperature profile were thoroughly tested and repeated in the lab to ensure a consistent operation and repeatable thermal profile for all subsequent experiments. Firstly, the sample was heated to ~810 °C and held for 15 min to homogenize the temperature. Then the melt was cooled with a cooling rate of 10 °C/min (~0.17 °C/s) until the melt was completely solidified (Fig. 1c).

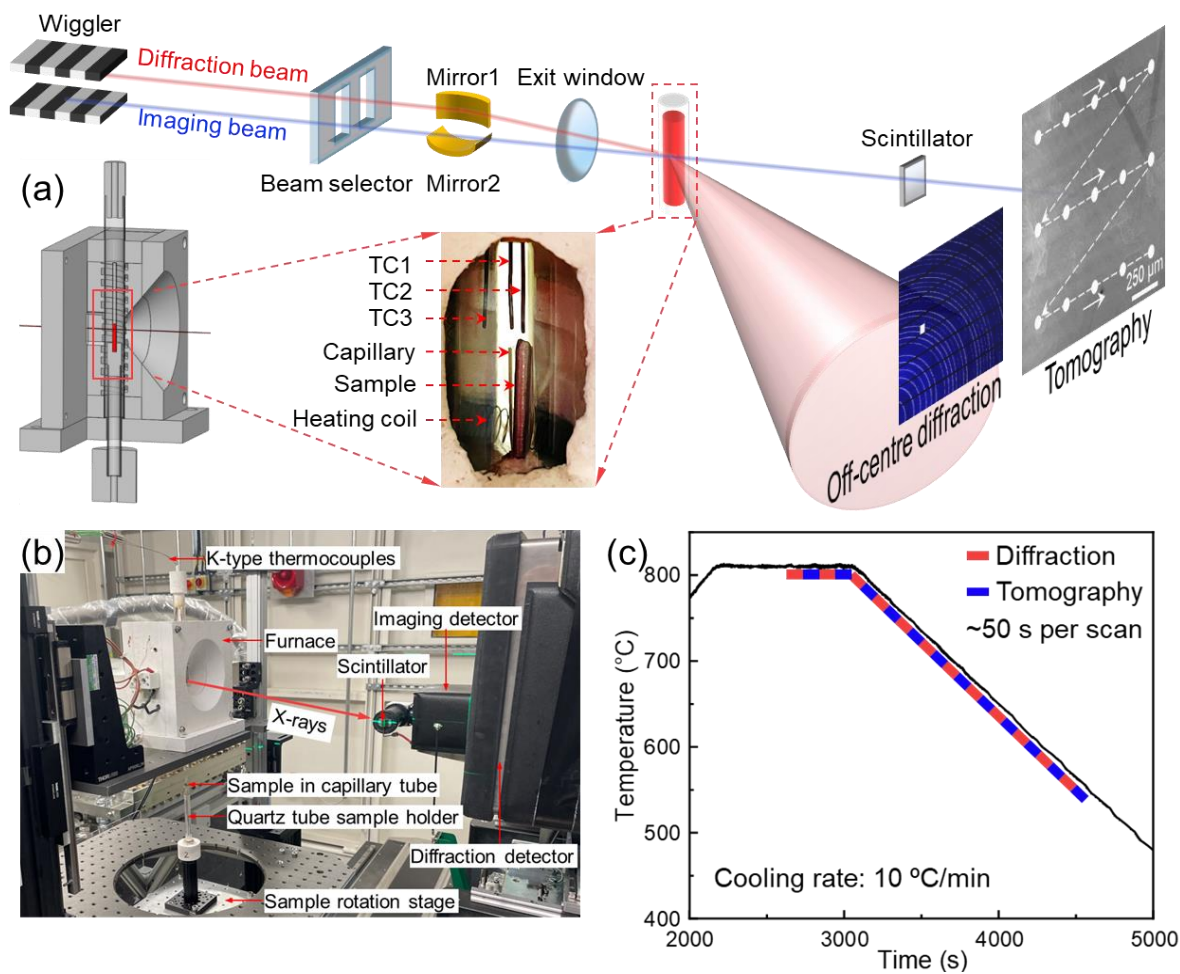

Fig. 1. (a) A schematic of the experimental set-up for the quasi-simultaneous synchrotron X-ray diffraction and tomography experiments at the DIAD beamline (K11) of the DLS. (b) A photo, showing the furnace and set-up on the sample stage at DIAD. (c) The heating and cooling temperature profile during the solidification process. Video 1 illustrates more clearly how the diffraction and tomography scans were executed.

## 2.3. The synchrotron X-ray parameters used

The beamline has an innovative optics layout that can "guide" two independent X-ray beams to reach the sample position in parallel for diffraction and imaging (tomography) acquisition [34]. A beam selector (see Fig. 1a) is positioned downstream of the wiggler for selecting either the diffraction or imaging beam for data acquisition, i.e., selecting one but blocking the other in order to avoid any crosstalk of the two X-ray signals. Switching between the two beams can be realised in a few Hz. Downstream of the beam selector, a pair of Pt-coated Kirkpatrick–Baez (KB) mirrors are used for 'moving' and focusing the diffraction beam onto the pre-designed spots of the sample (see Fig. 1a) [34]. However, there are no further optics for the imaging beam. For the diffraction scans, a monochromatic X-ray beam of 24.85 keV (25 × 25 µm spot size) and a PILATUS3 X CdTe 2M detector (Dectris AG, Baden-Daetwill, Switzerland) were used. The diffraction beam was from two horizontally reflecting mirrors and a horizontal bounce Si (111) monochromator. Diffraction patterns were collected on a grid of 5 × 3 points (each pattern with 3 s exposure time). The scan path was shown on the tomography projection in Fig. 1a. The step size was 250 µm in horizontal and 500 µm in vertical direction. The beam did not overlap between any two adjacent diffraction points. The scan was made by "moving" the KB-mirror instead of moving the sample [36]. A complete set of diffraction scan (i.e., 15 patterns, including dark-field and flat-field collection) took ~50 s.

Table 1. The experimental parameters used for diffraction and tomography acquisitions

| Diffraction (monochromatic beam) | | Tomography (pink beam) | |
|---|---|---|---|
| X-ray energy | 24.85 keV | X-ray energy | 20.00 keV |
| Beam size | 25 × 25 µm | Beam size | 1.7 × 1.7 mm |
| Detector | PILATUS3 X CdTe 2M | Camera | PCO.edge 5.5 |
| Pixel size | 172 × 172 µm | Effective pixel size | 0.54 µm |
| Image size | 1475 × 1679 pixels | Field of view | 2560 × 2160 pixel |
| Calibrant | $LaB_6$ 660b | No. of projections | 3000 |
| Exposure time | 3.0 s | Exposure time | 1.0 ms |
| Sample-to-detector distance | ~411 mm | Sample-to-scintillator distance | ~135 mm |

After each compete set of diffraction scan, a pink beam of 20 keV and a PCO.edge 5.5 camera (Excelitas Technologies, Waltham, MA 02451) were used for the tomography scan. The scintillator was coupled with a 12× optical microscope, achieving an effective pixel size of 0.54 µm in a field of view (FOV) of ~1.4 (width) × 1.2 (height) mm (2560 × 2160 pixels). Each full tomography scan (including dark-field, flat-field collection, and data transferring) also took ~50 s to complete. Video 1 shows a combination of the temperature profile, the diffraction and tomography scan path. Table 1 lists the parameters used for diffraction and tomography.

### 2.4. Data processing

The diffraction data were pre-processed using DAWN (v2.23.0) [37] and Fit2D (v18.002) [38], including masking out the defect pixels, subtracting background, calibrating the detector relative to the X-ray beam (in the yaw and pitch direction), azimuthal integration. The tomography data was pre-processed using the SAVU pipeline [39], including dark-field and flat-field correction, centre-of-rotation finding, and ring artefact removal and reconstruction. Phase index and identification in the 1D diffraction spectra were made using Match!3 [40]. 2D tomographic slices were post-processed using ImageJ (1.54f, NIH, USA) [41] and rendered using Avizo 3D (2021.1, Thermo Fish Scientific, USA).

## 3. Results
### 3.1. Phases and morphologies acquired quasi-simultaneously during solidification

Fig. 2a shows the scan path (direction) of the diffraction beam and the 15 points (i.e., 5 × 3 points) where diffraction patterns were collected inside the imaging FOV (i.e., the tomography projection). Fig. 2b presents the X-ray diffraction intensity spectra of 3 typical points (red, yellow, and green). Figs. 2c (top cross-sectional view) and 2d (front cross-sectional view) show the 2D tomographic slices collected immediately after each full diffraction scan in the solidification process. Fig.b1 shows that, at 768 °C (the black line), there is no crystalline phase peaks, indicating that the alloy melt was in a full liquid state. As the melt was cooled down, the tomography (Figs. 2c1 and 2d1 at 760 °C) collected right after the diffraction scan showed a polygonal structure appeared, which was identified as the $Al_4Mn$ phase (HCP structure, a=b=1.998 nm, c=24.673 nm, space group: P63/mmc, 194). As further cooling of the melt, more $Al_4Mn$ phases of different sizes appeared in the FOV (Figs. 2c2 to c4 and 2d2 to d4), and the diffraction peaks became more prominent, indicating that these $Al_4Mn$ phases were growing rapidly in the melt. Below the peritectic temperature (i.e., 705 °C, see Fig. S1 of the Supplementary Materials), more new and intensive diffraction peaks appeared (the purple spectrum shown in Figs. 2b1, b2 and b3), which were indexed as the $Al_6Mn$ phase (orthorhombic structure, a=0.755 nm, b=0.650 nm, c=0.887 nm, space group: Cmcm, 63). The immediately after tomography scan (Figs. 2c5 and 2d5 at 688 °C) shows that most of the early formed simple polygonal structures were transformed into a more complex cauliflower-type morphology, indicating peritectic transformation occurred at this stage. When the temperature was below 660 °C, two sharp peaks near $2\theta = 12.2°$ and 14.1° appeared, indicating the formation of the α-Al phase. After this, the diffraction spectra remained the similar profiles until 300 °C. Hence, the fully solidified sample consists of α-Al, $Al_4Mn$ and $Al_6Mn$ phases. The diffraction results (Fig. 2b) acquired along the scan path show that the diffraction peaks are different in positions and intensity, indicating that the nucleation, growth and transition of the phases occurred independently at different locations within the melt volume. There was no directional growth behaviour found in the solidification process, which was in consistent with the observation of the phases' spatial distribution in 3D from the tomography (Figs. 2c, 2d and Fig.3). The diffraction patterns in the other 12 points have the similar characteristics and therefore are not repeated here.

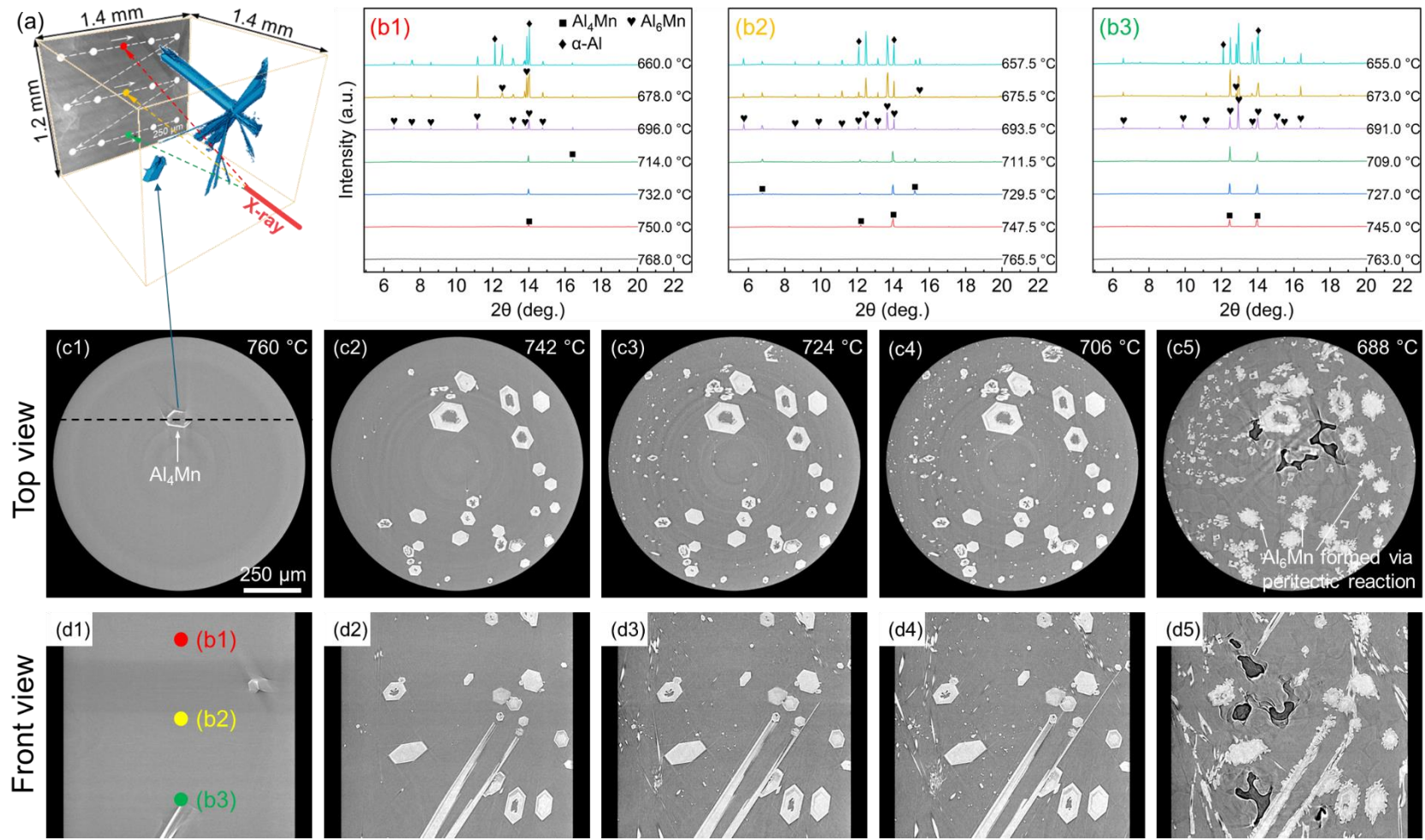

Fig. 2. (a) The scan path (direction) of the diffraction beam inside the imaging FOV. (b1) to (b3) Three diffraction intensity spectra of 3 typical points in Figs. 2a (more clearly marked in d1). (c1) to (c5) The top cross-sectional views of typical tomography scans collected immediately after each full set of diffraction scan (i.e., 5 × 3 points) during the solidification process. (d1) to (d5) The front cross-sectional views according to the black dashed line in Fig. 2c1. Please see Video 1 for the dynamic evolution.

Figs. 3a to 3f show the 3D rendered results of the Al-Mn intermetallics in the entire scanned volume. Although the 2D tomography slices show these intermetallics as individual polygonal structures, in 3D space, they were actually long rod-shaped structures impinged or crossed over each other during solidification, forming extremely complex truss structures. Fig. 3g shows the phase volume fraction as a function of temperature measured from the 3D datasets. The calculated volume fraction using the Scheil-Gulliver model in JMatPro® (v13.2) is also shown for comparison [42]. The measured fractions were in agreement with the Scheil-Gulliver model in the early stage of solidification (760-710 °C) but deviated more in the later stage. In the following sections, we extracted some representative cases (typical morphologies) from the whole volume to discuss and elaborate on the phase growth and transition dynamics.

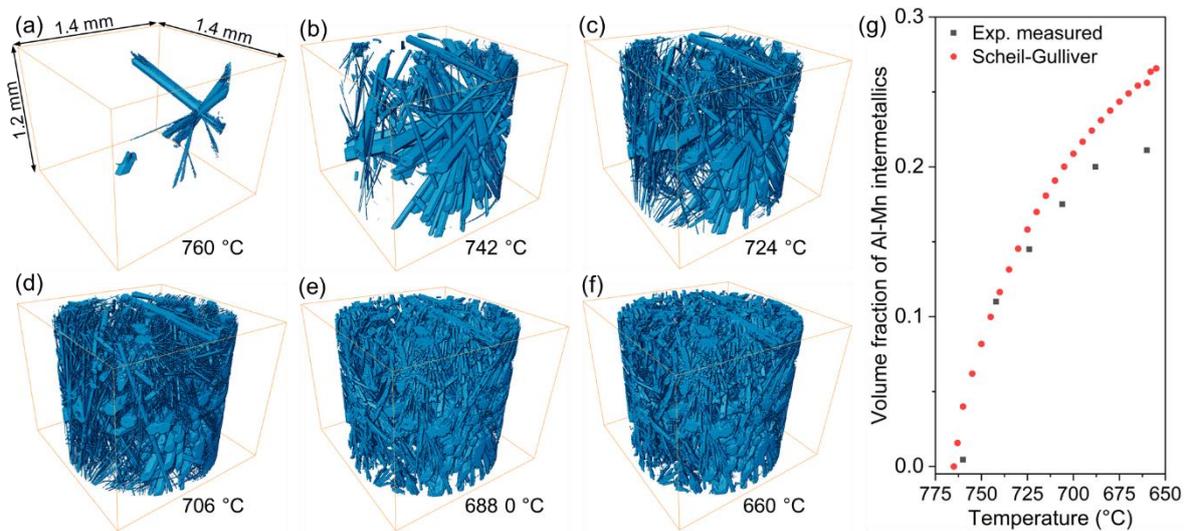

Fig. 3. (a) to (f) the Al-Mn intermetallics rendered from the acquired tomography data, showing the total phase volume as a function of temperature . (g) The measured Al-Mn phase volume fraction versus that calculated by using the Scheil-Gulliver model in JMatPro® (v13.2).

### 3.2. Growth dynamics of the primary $Al_4Mn$ phase and formation of core defects

Fig. 4 shows the morphological evolution in 3D for 3 different but typical $Al_4Mn$ phases (named V1, V2 and V3) and the formation of core defects in solidification from 760 °C to 706 °C (i.e., prior to the peritectic reaction). For case V1, Fig. 4a and Video 2 show that a rod-shaped long $Al_4Mn$ phase (8.1 µm in diameter, 513 µm in length) was seen to appear in the FOV. Firstly, it grew rapidly along the axial direction (the vertical direction) with an average speed of 3.9 µm/s (Figs. 4d to f) until its growing front was intercepted and obstructed by an $Al_4Mn$ phase that grew in the horizontal direction (see the phase marked by H1 in the inset of Fig. 4a). Such obstruction forced V1's growing front to split into two smaller branches, which continued to grow in the same direction until once again each of them was impeded by H2, then H3, etc.

each such obstruction forced more split and therefore more branches (see the top inset in Fig. 4a). At the same time, V1 also grew radially, but with a much slower speed of only 0.056 μm/s (~70 times slower than the axial growth speed, see Fig. 4f). Fig. 4b and Video 3 show the case V2. Initially, it exhibited a typical hexagonal prism shape (~40 μm in width, ~395 μm in length). It grew in the axial direction (2.6 μm/s) and radial direction (0.17 μm/s) without any obstruction (Fig. 4f). However, a number of core defects developed at the growing front, and gradually such defects grew bigger and bigger and then merged to form large tubular core structures (see the top insets in Fig. 4b).

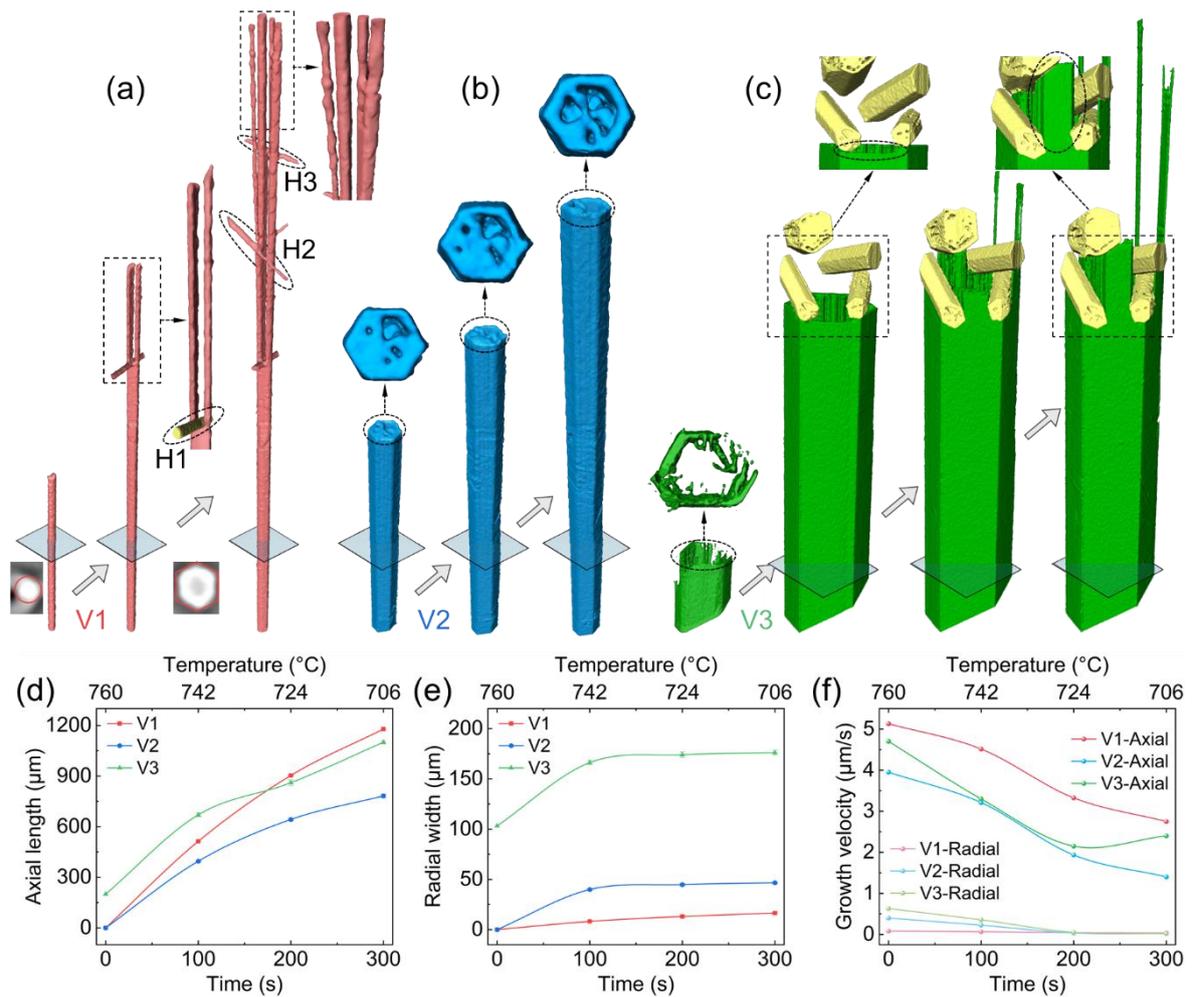

Fig. 4. (a) to (c) Morphological evolution in 3D for 3 typical $Al_4Mn$ phases of different sizes (named V1, V2 and V3, respectively) during solidification, showing the growth dynamics in the axial direction and radial direction respectively (Videos 2, 3 and 4 show more vividly the growth dynamics). (d) The axial length and (e) radial width growth of the 3 phases as a function of time and temperature. (f) Growth velocity of the 3 phases in the axial and radial direction.

Fig. 4c and Video 4 show the case V3. It started with an incomplete hexagonal prism shape (~200 μm in length, ~103 μm in width, first appeared in the FOV at 760 °C) with a distinct core defect inside the hexagon. Again, due to the obstructions by the phases growing from other directions, the top surface of the prism split into several platelets that continued to grow axially like the behaviour of V1 in Fig. 4a. The average growth speed of V3 was ~3.1 μm/s in the axial direction and ~0.26 μm/s in the radial directions respectively (Fig. 4f).

### 3.3. Solute diffusion and redistribution during growth of the primary Al$_4$Mn phases

To understand the effect of solute Mn diffusion and distribution on the phase growth dynamics, we extracted the Mn solute distribution information based on the variation of the grey values in the tomographic slices. The correlation and calibration were established using the composition line scan from the energy dispersive X-ray spectroscopy (EDXS) measurements as detailed in the sections 2-4 of the Supplementary Materials. Such calibration established a linear relationship between the grey value (G) and Mn concentration as:

$$C_{Mn} = 0.133\,G - 8.296 \qquad (1)$$

which was then applied to the tomography slices to map the Mn distribution surrounding the growing Al$_4$Mn phases (Figs. 5a to c). The solute Mn distribution information is critical for understanding the subsequent peritectic reaction and phase growth dynamics.

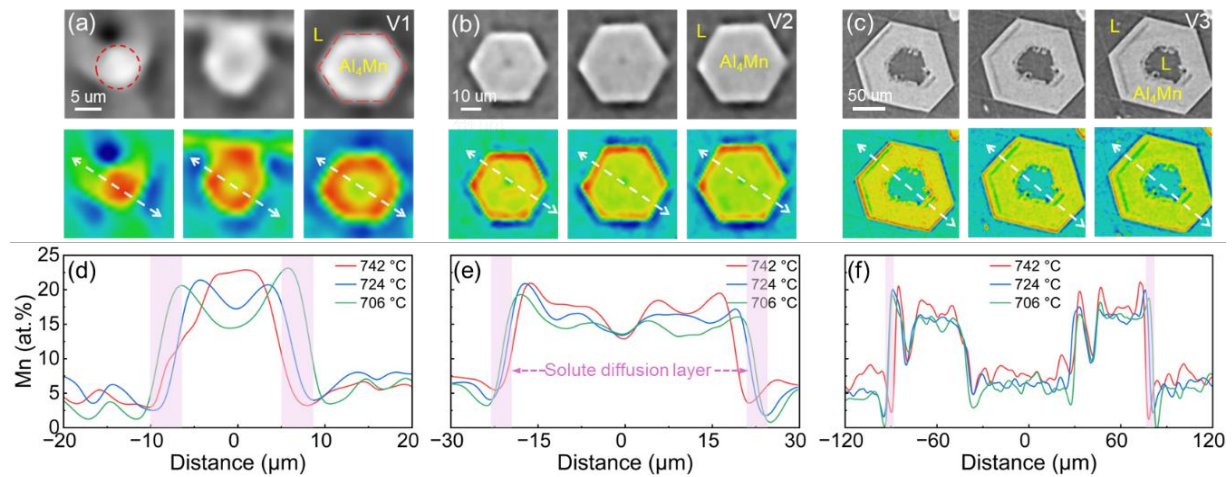

Fig. 5. In the 1$^{st}$ row of (a) to (c): typical tomography 2D slices of V1, V2 and V3 at the cross-sections indicated by the shaded squares in Figs. 4a to c. In the 2$^{nd}$ row of (a) to (c): the corresponding pseudo-colour X-ray attenuation maps. (d) to (f) The Mn distribution profiles along the white dash lines in the pseudo-colour maps. The phase centre is set as zero position.

Figs. 5d to f show the Mn distribution profiles across the liquid-solid-liquid interface of the three representative Al$_4$Mn phases (i.e., V1, V2, V3 in Fig. 4). In the liquid region (the green colour region) away from the liquid-solid interface, the Mn concentration maintained at ~5.0 at.% with relatively flat profiles for all 3 cases, followed by a pronounced Mn-depleted zone in the liquid region (the blue colour region) near the interface. However, moving into the solid region (the red and then yellow colour region), there were sharp increases in Mn concentrations at the outer edges of the solidified $Al_4Mn$ phase. Such steep Mn concentration changes were constrained within the liquid-to-solid transition zone with its thickness measured at ~5 μm (as indicated by the purple shaded area in Figs. 5d, e and f), which are very consistent for all 3 cases. The formation of such Mn-enriched zone in front of the solid-liquid interface also simultaneously created a solute-depleted zone in the immediate liquid region. As solidification proceeded and temperature decreased, Mn depletion in the immediate liquid region became more prominent (Figs. 5d to f). Clearly, this thin Mn diffusion zone is controlling the growth dynamics of the $Al_4Mn$ phases from the liquid region.

### 3.4. Peritectic reaction dynamics ($Al_4Mn$ + liquid → $Al_6Mn$) and phase spatial relationships

According to the Al-Mn phase diagram (Fig. S1), the peritectic reaction ($Al_4Mn$ + liquid → $Al_6Mn$) occurred at ~705 °C (Fig. 6a). Figs. 6b and c show the typical primary $Al_4Mn$ phase. Figs. 6d and e show the resulting peritectic phase ($Al_6Mn$). The $Al_6Mn$ phases grew into irregular and rugged strip-shaped structures on the interior and exterior facets of the $Al_4Mn$ phases, forming a complex cauliflower-shaped morphology. Video 5 shows more vividly the peritectic reaction dynamics. In fact, the spatial relationship between the $Al_6Mn$ and the $Al_4Mn$ in 3D view (Figs. 6d and e) is far more complicated than that illustrated by the 2D sectional views (compared the sectional views shown by Figs. 6f1 and f2 with the corresponding real 3D structures shown by Figs. 6g1 to g4).

In order to determine the spatial relationships between the $Al_4Mn$ and $Al_6Mn$, Figs. 6f and g present a rendered peritectic structure in 3D along with two 2D slices in the chosen locations. A thin shell structure in blue colour tightly wrapped around the reddish precursors. Such thin shell was the direct product of the peritectic reaction (the region marked by the red lines in Figs. 6f1 and f2). In fact, in a non-equilibrium solidification condition, the $Al_4Mn$ was not able to fully transform into $Al_6Mn$. The non-uniform solute distribution and phase impingement resulted in an irregular outer surface morphology for the $Al_6Mn$ phases. In addition, there existed some prism-shaped structures that grew epitaxially into the melts (see the blue line circulated areas in Figs. 6f1 and f2). 3D view indicates clearly that these structures also have pronounced anisotropic growth behaviours (Figs. 6g1 to g4), i.e., they grew more quickly along one particular direction and eventually formed long prism-shaped structures. Interestingly, at the centres of these $Al_6Mn$ phases, there were also distinct regions filled with α-Al phases and these regions grew bigger and bigger as the phase grew (Figs. 6g2 and g4), similar to those observed in case V2 of Fig. 4b.

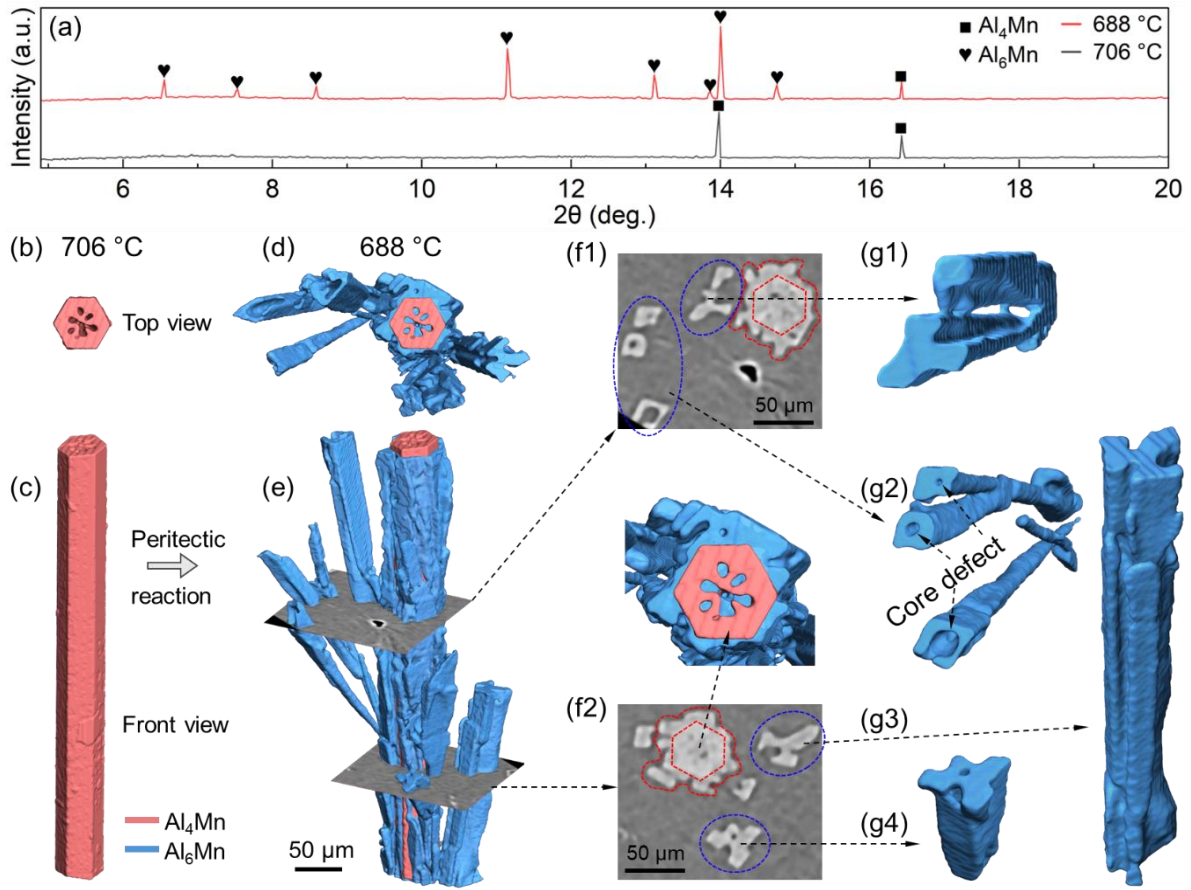

Fig. 6. (a) The synchrotron X-ray diffraction spectra, showing the phase transformation. (b) and (c) 3D rendered views of a typical peritectic precursor (Al$_4$Mn). (d) and (e) The resulting peritectic structure in 3D along with two 2D planes at the chosen locations, showing the spatial relationship between the peritectic precursor (Al$_4$Mn) and the resulting product (Al$_6$Mn). (f1) and (f2) The enlarged 2D tomography slices corresponding to the two planes shown in Fig. 6e. (g1) to (g4) 3D rendered results of the prism-shaped Al$_6$Mn phases growing epitaxially into the melts, showing a pronounced anisotropic growth behaviour with core defects formed at the centre.

### 3.5. Crystallographic orientation relationships between the peritectic precursors and products

Previous research showed that the crystallographic orientations of the peritectic products are definitely related to the orientation of the precursors [7, 43]. Here, we used electron backscattered diffraction (EBSD) coupled with EDXS to further analyse the coupled peritectic structures of the fully solidified alloy sample, as detailed in the section 5 of the Supplementary Materials. The overlaid phase map and elemental distribution maps (Figs. 7a to d) confirm that the inner region (red) was HCP Al$_4$Mn phase with lower Al and a higher Mn, while the outer region (blue) was orthorhombic Al$_6$Mn phase (see Table 2).

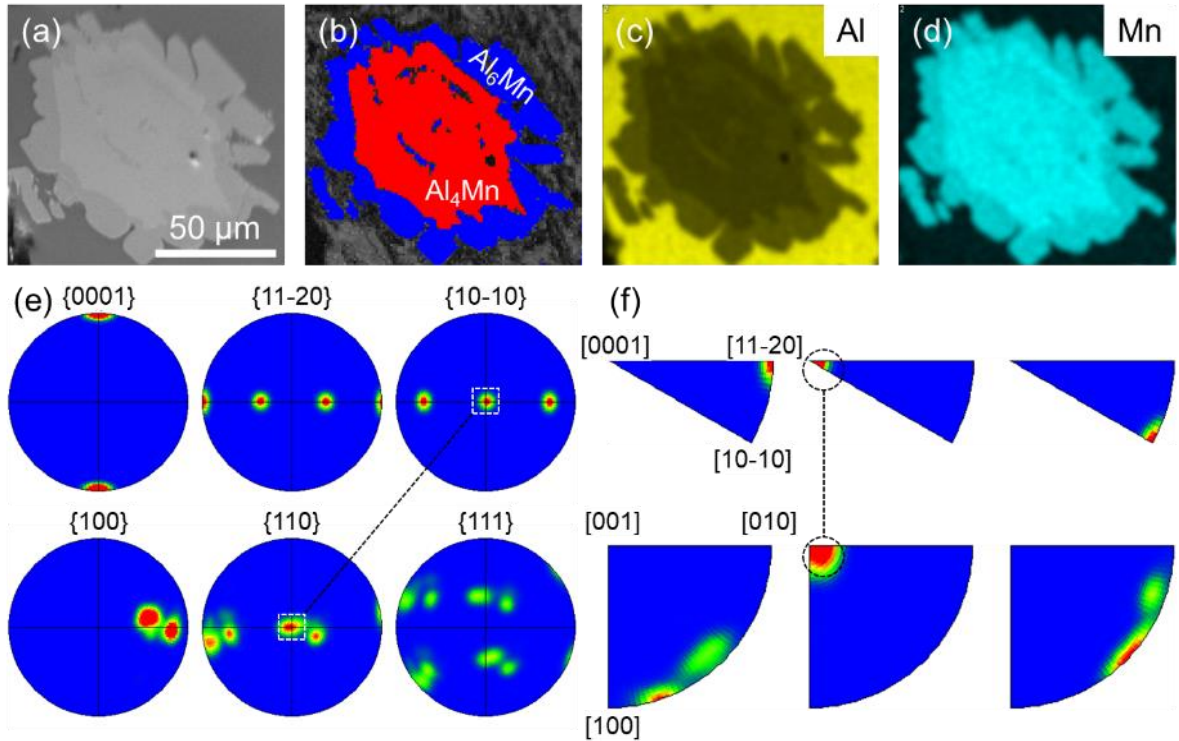

Fig. 7. (a) SEM image, showing the coupled peritectic structures of the fully solidified sample. (b) Phase map overlaid with band contrast map. The red part is indexed as $Al_4Mn$ phases, and the blue part is indexed as $Al_6Mn$ phase. (c) and (d) EDXS element map of Al and Mn. (e) and (f) Pole figures and inverse pole figures of the two phases.

Table 2. The chemical compositions of the $Al_4Mn$ and $Al_6Mn$ phases measured by EDXS.

| Phase | Al (at.%) | Mn (at.%) |
|---|---|---|
| $Al_4Mn$ | 81.42 ± 0.02 | 18.58 ± 0.04 |
| $Al_6Mn$ | 85.88 ± 0.05 | 14.12 ± 0.05 |

We also analysed and determined the orientations of the coupled peritectic structures. In order to clearly elucidate the orientation relationships, the coordinate system of the samples was adjusted in the virtual chamber to allow the basal plane (i.e., {0001}) of the $Al_4Mn$ phase to be horizontally perpendicular to the screen. Figs. 7e and f show the adjusted pole figures and inverse pole figures of the $Al_4Mn$ (top) and the $Al_6Mn$ (bottom) phases, respectively. By comparison, we can clearly see that the pole of the $Al_6Mn$ phases is mostly located at the centre of the {110} pole figure, which just corresponds to the centre of the {10-10} polar figure of the $Al_4Mn$ phases (as indicated by the white rectangles in Fig. 7e), and also the pole in the [001] direction on its inverse pole figures likewise corresponds to that in the [0001] direction of the $Al_4Mn$ phase (as indicated by the black circles in Fig. 7f). The highly overlapping poles undoubtedly demonstrate the existence of a specific crystallographic orientation relationship between the peritectic precursor (HCP Al$_4$Mn phase) and its product (orthorhombic Al$_6$Mn phase), i.e., {10-10}$_{HCP}$ // {110}$_O$, [0001]$_{HCP}$ // [001]$_O$.

## 4. Discussion

### 4.1. Nucleation and anisotropic growth dynamics of the primary Al$_4$Mn phases

Our previous studies have shown that solute and solvent atoms in Al alloy melts gradually rearrange to formed polyhedral atomic clusters of a certain degree of crystalline-type order by overcoming the lower solid-liquid interfacial energy barrier during solidification [28, 44]. These polyhedral atomic clusters with higher degree of crystalline symmetry are proved to be the nucleation "precursors" for the mature crystals in the liquid-solid coexisting region [28, 35]. According to the classic nucleation theory [45], once the nucleation "precursors" overcome the free-energy barrier at a critical nucleus size, they would grow out to form mature crystallites in the heterogeneous liquid (Fig. 8a1). Following on nucleation, the newly formed crystallites move onto the growth stage, where they mature into smooth faceted macroscopic or rough non-faceted dendritic crystals [46].

Based on our diffraction (Fig. 2b) and tomography results (Figs. 2c1 and d1), the Al$_4$Mn phases are crystallized primarily from the melt when the temperature drops below the liquidus temperature (~760 °C for Al-8wt.%Mn alloy), followed by the formation of Al$_6$Mn phases via a peritectic reaction at ~705 °C. The tomography results demonstrate that both Al$_4$Mn and Al$_6$Mn phases grow into elongated prism-shaped structures enclosed by smooth facets (Figs. 4 and 6), demonstrating a pronounced anisotropic faceted growth behaviour. Theoretically, an ideal bulk crystal may possess an infinite number of exposed faces. However, the actual grown crystals possess a limited number of exposed faces [47]. In general, the morphology of a growing crystal is mainly determined by the intrinsic crystallographic kinetics and the external thermodynamic conditions [17]. The former affects the growth dynamics along certain crystallographic planes and directions, in particular for the crystals with high anisotropy.

Table 3 lists the interplanar distances of the most common crystal planes of Al$_4$Mn and Al$_6$Mn phases. Al$_4$Mn phase with an HCP crystal structure has the largest interplanar distance of 2.4673 nm along the basal plane (0001) and the second with 1.7303 nm along the prismatic plane (10$\bar{1}$0). Therefore, the <0001> direction (i.e., along the c-axis, perpendicular to the basal plane) is relatively loosely packed, with weaker atomic bonding between layers, hence having lower energy barrier for atomic attachment. In contrast, the <10$\bar{1}$0> direction (i.e., along the a-axis, perpendicular to the prismatic plane) is more tightly packed, with stronger in-plane atomic coordination, leading to a higher kinetic barrier for growth. Consequently, atoms (or atomic clusters) from the melt attach more easily and grow along the <0001> direction, resulting in a much faster intrinsic growth rate compared to the more constrained <10$\bar{1}$0> direction (see Figs. 8a2 and a3).

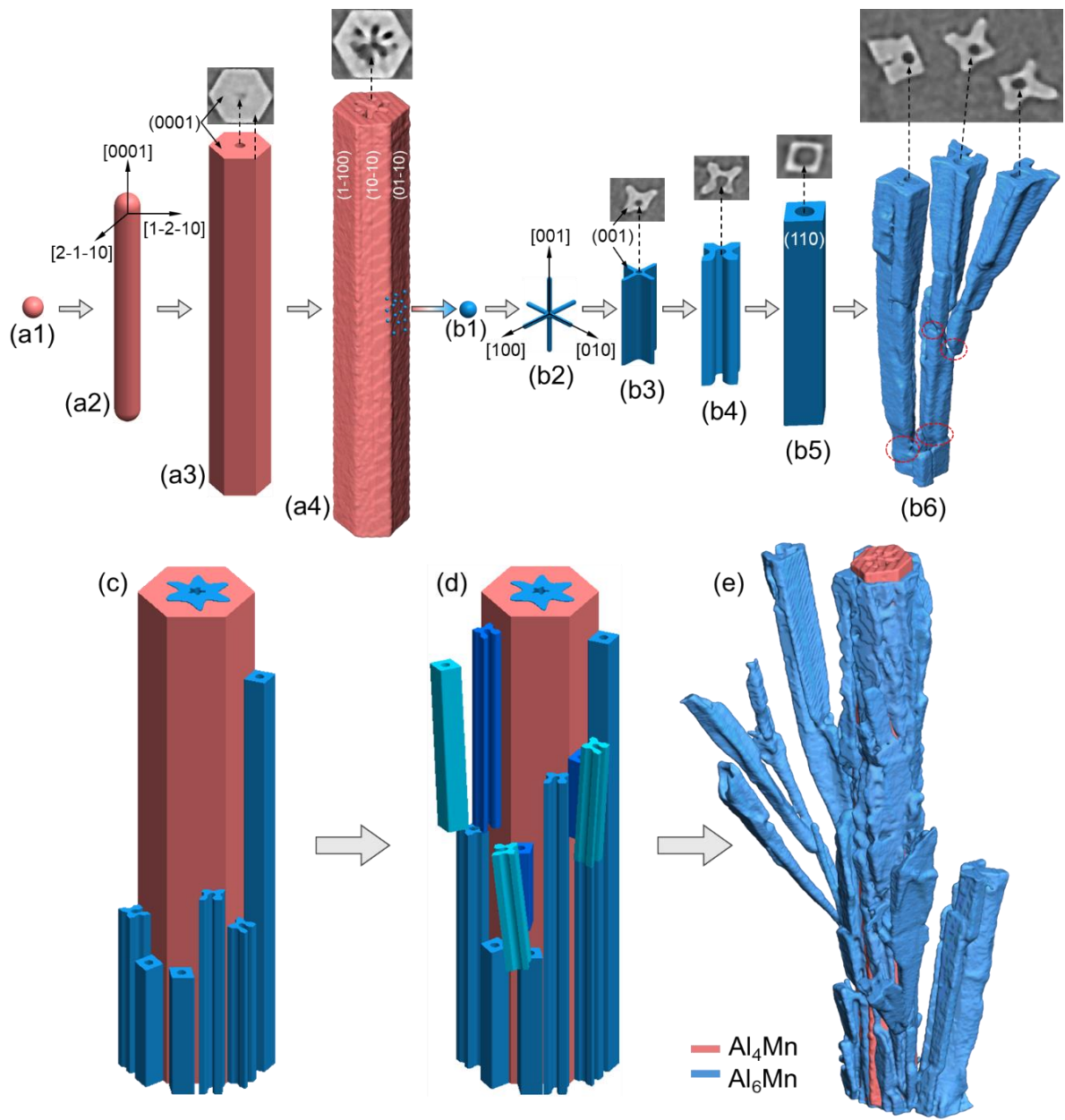

Fig. 8. Schematic representation of the nucleation and growth mechanisms: (a1) to (a4) for the Al$_4$Mn phases, (b1) to (b6) for the Al$_6$Mn phases. (c) and (d) The evolutionary mechanism of the peritectic reaction. (e) A representative peritectic structure extracted from the volume, showing the intertangled truss morphology in the 3D view. Please also see Video 5 for more detailed dynamics.

Furthermore, in accordance with the Wulff construction theorem [48], a growing crystal always requires minimizing its total surface energy for a given crystal volume [47]. In most intermetallic growth cases, a growing crystal would be bounded by the faces with the largest interplanar distances and lowest surface energies, eventually resulting in the formation of anisotropic faceted morphology [47, 49, 50]. To minimize the total energy, the $Al_4Mn$ crystal therefore grows preferentially along the <0001> direction, reducing the area of the higher-energy prismatic faces. This directional preference leads to the development of elongated prismatic morphologies with high aspect ratio, as observed in Figs. 4 and 8a4.

Table 3. The interplanar distances of some crystal planes of $Al_4Mn$ and $Al_6Mn$ phases [51].

| Planes in $Al_4Mn$ | Interplanar distance/nm | Planes in $Al_6Mn$ | Interplanar distance /nm |
|---|---|---|---|
| (0001) | 2.4673 | (001) | 0.8870 |
| ($10\bar{1}0$) | 1.7303 | (100) | 0.7550 |
| ($10\bar{1}1$) | 1.4167 | (010) | 0.6500 |
| ($10\bar{1}2$) | 1.0045 | (110) | 0.4926 |
| ($11\bar{2}0$) | 0.9990 | (111) | 0.4306 |

In solidification, anisotropic crystal growth is also influenced by the thermodynamic conditions, including solute diffusion, heat transfer, cooling rate, and solute rejected fields caused by the competitive growth between adjacent crystals [25]. More importantly, our experiments revealed that there exists a thin Mn-rich diffusion boundary layer (~5 μm thick, with a deeper depletion in the liquid side and higher supersaturation in the solid side, see Fig. 5). It establishes a local solute gradient that controls and dominates the growth kinetics at the liquid-solid interface, making the $Al_4Mn$ growth dynamics have a typical diffusion-controlled characteristics. Clearly, in the radial direction (i.e., <$10\bar{1}0$>), accumulation of the Mn atoms within this thin layer creates a barrier to slow down (or stop) any further solute diffusion in the radial direction, hence very slow growth in the radial direction initially and then completely terminated. Conversely, along the fastest-growing axial direction (i.e., <0001>), the high growth velocity prevents the establishment of a similar solute boundary layer at the tip, allowing the crystal to outrun the solute-depleted region and maintain sustained longitudinal growth (see Figs. 4 to 6). Such diffusion-controlled growth dynamics also explained well the observed discrepancy between the experimentally measured phase volume fraction and that calculated by using the Scheil-Gulliver model (Fig. 3g). The Scheil model assumes a complete and homogeneous solute diffusion in the liquid state. However, in our experiments, the pronounced Mn-depleted zone in the liquid side at the solid-liquid interface effectively changes the local solute concentration, suppressing the nucleation probability in the liquid region further away from the growing phases, thus effectively reducing the total phase volume fraction at the later stages of solidification.

### 4.2. Nucleation and growth dynamics of the peritectic $Al_6Mn$ phases

### 4.2.1. Nucleation and anisotropic growth

The Al$_6$Mn phase has an orthorhombic structure and the interplanar distance of the (001), (100), and (010) plane is 0.887 nm, 0.755 nm, and 0.650 nm, respectively. Therefore, the growth rate along the direction [i.e., <001> of (001)] plane will be greater than that on the (100) and (010) planes (Figs. 8b2 and b3). Similarly, due to crystal anisotropy, the Al$_6$Mn crystal continued to grow radially along the direction [i.e., <110> of the (110)] plane in order to reduce the total surface energy of the growing crystal (Figs. 8b4 and b5), developing into a tetragonal prism-shaped morphology.

### 4.2.2. Re-nucleation and non-anisotropic branching

In addition, at the edges and/or corners of the growing phase fronts during the growth of the Al$_6$Mn phases, re-nucleation and phase branching are often found to occur as highlighted by the red circles in Fig. 8b6. The branching dynamics are driven by the competition between thermodynamic instability and crystallographic kinetics. From thermodynamic point of view, the flat and smooth crystal facets have the lowest surface energy and are thus thermodynamically more stable. However, the edges and corners are formed due to two or multiple crystal planes intersection, and are the sites of high curvature and higher interfacial energy, making them energetically unfavourable [52]. To reduce the interfacial energy at those sites, these high-energy sites become preferential locations for phase re-nucleation. The newly nucleated crystals may also change their crystallographic orientation, either align coherently or semi-coherently with that of the existing phase in order to promote crystallographic continuity and therefore reduce the surface free energy at the junctions [52]. From the kinetic point of view, the incomplete progress of peritectic reaction and the preferential growth of Al$_6$Mn crystals lead to local solute redistribution in the melt, particularly, the Mn concentration is consumed largely near the growth front, creating a solute-depleted region ahead of the liquid side of the solid-liquid interface and a concentration gradient near the corners and edges [17]. These areas experience significant solute gradient, increasing the driving force for nucleation of new Al$_6$Mn crystals. Once a new crystal nucleates at the high-energy sites (corners or edges), it begins to grow along its favourable crystallographic directions, which in turn changes the local solute concentration nearby, potentially triggering other re-nucleation events closely. Thus, the Al$_6$Mn growth front undergoes repetitive cycles of local solute gradient, re-nucleation, re-growth, and termination. Eventually, the Al$_6$Mn phases grow into complex dendritic or multi-segmented anisotropic structure or morphology as illustrated in Figs. 6 and 8b6.

### 4.3. Growth dynamics of the peritectic phases at a low cooling rate

Through the above analyses, we have elaborated clearly the nucleation and growth mechanisms of the primary $Al_4Mn$ phases and peritectic $Al_6Mn$ phases, as well as their compositional, spatial and orientation relationships. From these new findings, we can reveal clearly the growth dynamics of the peritectic phases at the cooling rate of 10 °C/min (~0.17 °C/s), as illustrated in Figs. 8c to e and Video 5.

Firstly, when a primary $Al_4Mn$ phase is formed and grows into a certain size, a stable Mn-enriched diffusion boundary layer (~5 μm thick) is created at the solid-liquid interface (see Fig. 5). This local region, with the Mn concentration significantly higher than that in the bulk liquid, plays the dominant role in controlling the subsequent peritectic phase transformation. At the peritectic temperature, nucleation of $Al_6Mn$ on the $Al_4Mn$ surfaces occurs in direct contact with the melt (Fig. 8a4). Theoretically, nucleation would occur anywhere on the solid-liquid interface [44]. However, the established Mn-enriched boundary layer provides the favourable surface for nucleation [44]. Crucially, such direct nucleation on top of the $Al_4Mn$ phase surface enables the newly formed $Al_6Mn$ phase to follow the specific orientation relationship: $\{10\bar{1}0\}_{HCP}$ // $\{110\}_O$, $[0001]_{HCP}$ // $[001]_O$ (see Fig. 8c). This low-energy and crystallographic orientation coupling allows the direct peritectic products to form a thin, continuous shell structure surrounding the primary $Al_4Mn$ phase (Figs. 6b1 and b2).

The subsequent growth of the peritectic $Al_6Mn$ phases is critically constrained by the geometry and solute mass of this boundary layer. The initial shell morphology is a direct consequence of the thin Mn-rich layer. As the locally rich Mn in the boundary layer is rapidly consumed, further growth of the $Al_6Mn$ phases requires long-range diffusion of Mn from the solute-depleted bulk liquid to arrive at the growing front of the newly formed $Al_6Mn$ shells. Such Mn diffusion is impossible to occur along the radial direction once a complete $Al_6Mn$ shell is formed, hence the growth in the radial direction stops after a $Al_6Mn$ shell is formed (see Figs. 4e and f). while in the axial direction, Mn depletion in the remaining melt would slow down the Mn diffusion, hence the phase growth in axial direction also slows down (see Figs. 4d and f). Such diffusion-controlled process renders, in most situation, an incomplete peritectic reaction. Further melt cooling provided more driving force, promoting more re-nucleation events to occur at the energetically favourable edges and corners of the hindered growing fronts (Fig. 8d). These re-nucleated crystals will grow epitaxially into the melt, forming branched structures. Consequently, the final microstructures exhibit intertangled truss morphology in 3D (Fig. 8e).

### 4.4. Nucleation and growth dynamics of the core defects

Another interesting phenomenon is that the formation of core defects within both the $Al_4Mn$ and $Al_6Mn$ phases during their growth process. These defects emerge along the growth axis and evolve into tubular core structures as shown more clearly in Fig. 9.

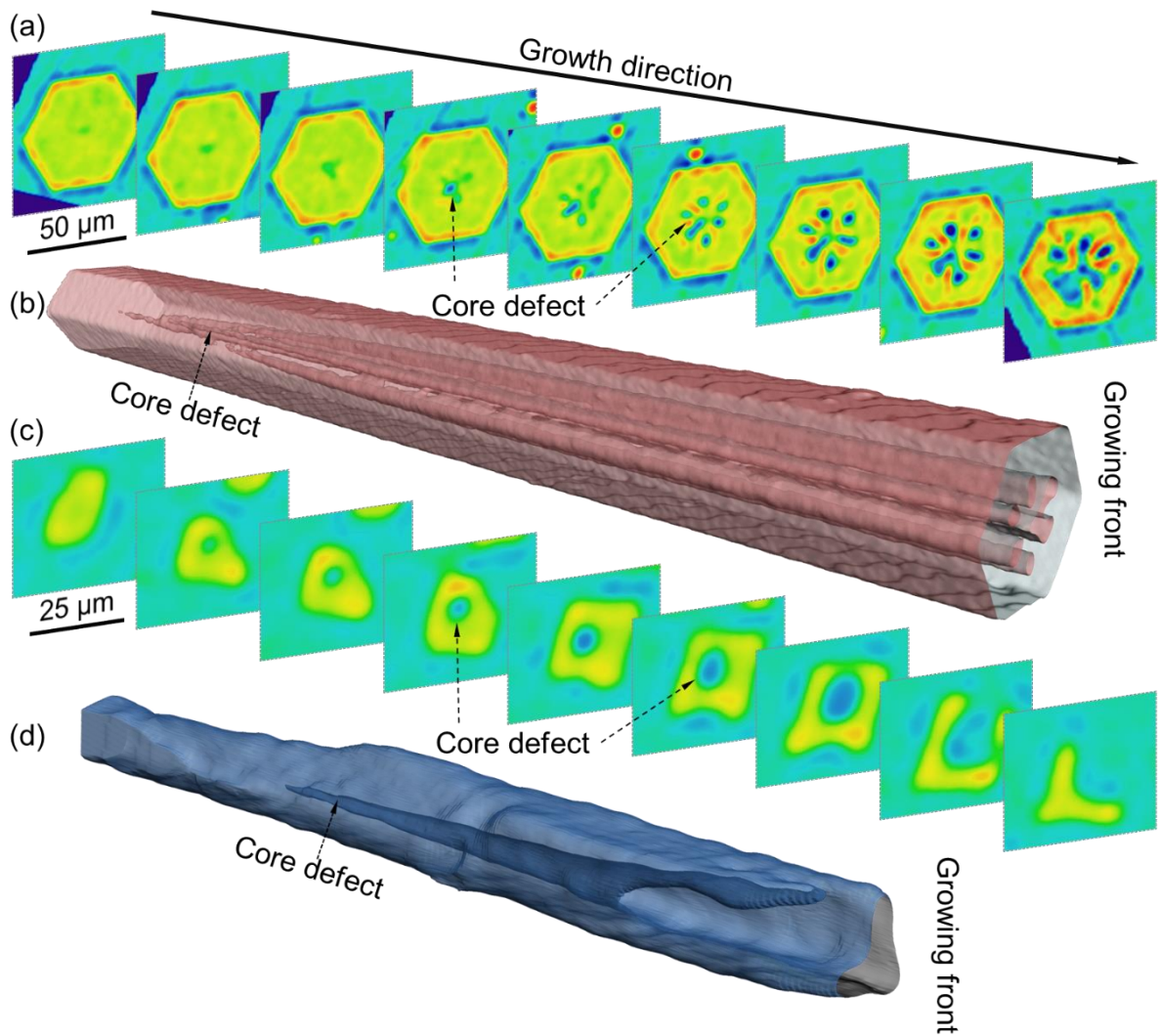

Fig. 9. Typical tomography and slices, illustrating the formation and evolution of core defects at the phase centre (α-Al phases are removed to highlight the morphologies of the Al-Mn intermetallic phases and defects). (a) A series of 2D slices of the $Al_4Mn$ phase along the growth direction and (b) the corresponding phase in 3D. (c) 2D slices of the $Al_6Mn$ phase and (d) the corresponding phase in 3D. The regions with lower attenuation are shown in blue and green, which normally mean a higher concentration of light elements or void (i.e. Al or hole). While yellow to red regions contains heavier elements (i.e., Mn), where the X-ray attenuation was strong [53, 54].

During the solidification process, both $Al_4Mn$ and $Al_6Mn$ phases exhibit faceted and uniaxial growth, primarily along their preferential crystallographic directions (Figs. 9b and d). However, due to the anisotropic diffusion, i.e., the long-range solute transport along the growing axis is significantly slower than the lateral feeding from the melt. Hence, the solute (Mn) supply at the centre of the growing front tip becomes more and more depleted as the phase continued to grow longer. Meanwhile, when the growing fronts encounter spatial obstructions or neighbouring crystals (e.g., the H1 - H3 phases in Fig. 4a), this locally mechanical impingement creates significant strain mismatch across the crystal facets, disrupting the ordered atomic stacking specifically at the central region where the solute flux is already restricted [55, 56]. In addition, the rapid crystallisation along the growing direction release significant latent heat at the liquid-solid interface. The latent heat at the outer surface may be dissipated into the melt, but the latent heat released at the inner surface (see Figs. 4c and 5c) may not be able to be dissipated easily, instead would act as the extra heat energy to remelt the trapped Al inside the core [56, 57], creating tiny liquid melt-filled channels along the crystal's growth axis. As solidification proceeds, the channel expands longitudinally along the growth directions and finally evolves into large and irregular channels that may contain some void if no sufficient liquid to back-fill that region (see Fig. 9). The channels actually co-grow with the phases. Similar phenomena have been reported in the anisotropic growth of various intermetallic phases, such as $Al_2Cu$ [58], $Al_3Sc$ [59], $Al_3Ni$ [56], and $Cu_6Sn_5$ [57], etc. What we have found here is that the nucleation and growth of the axial channels occur in both primary $Al_4Mn$ and peritectic $Al_6Mn$ phases, indicating that it is driven by the growth dynamics instead of nucleation and independent on the type of phase reaction. More importantly, unlike most previous studies based on post-mortem 2D microscopy characterisation, our studies are carried out in real-time and in 3D space, avoiding any ambiguity.

## 4.5. Effect of cooling rate on the growth dynamics and 3D structures of the peritectic phases

Cooling rate has profound influence on the thermodynamic and kinetic pathway of solidification and the final microstructures [14, 60]. Fig. 10 and Video 6 show the representative 3D rendered views along with 2D tomography slices at chosen locations for the samples solidified at cooling rates of ~0.17 °C/s (Fig. 10a), ~2.0 °C/s (Fig. 10b), and ~20 °C/s (Fig. 10c). The evolution of microstructures across these conditions vividly illustrates a transition from fully faceted to partly faceted and finally to non-faceted growth behaviour.

At the slow cooling rate of ~0.17 °C/s (Fig. 10a), the system is kinetically permitted to evolve towards its thermodynamic minimum. The $Al_6Mn$ phases develop into large and well-defined tetragonal prisms with high aspect ratios, as sufficient time allows for the atomic rearrangement and interface migration governed by anisotropic interfacial energy. The pronounced and continuous tubular channels are evident along the growth axes (as shown in the accompanying 2D slices in Fig. 10a), a signature of sustained anisotropic solute diffusion, impurity and lattice mismatch-induced strain accumulation/relaxation, as detailed in section 4.3.

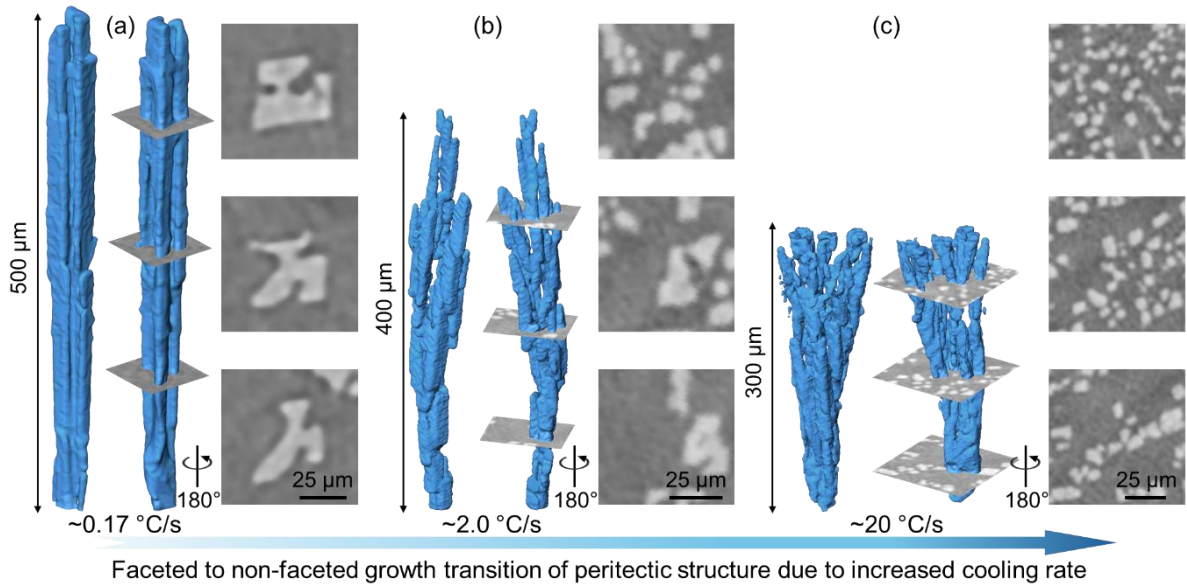

Fig. 10. (a–c) 3D reconstructions and corresponding 2D tomography slices of typical peritectic structures from Al-8wt%Mn alloys solidified at cooling rates of: (a) ~0.17 °C/s, (b) ~2.0 °C/s, and (c) ~20 °C/s. See Video 6 for appreciating the true 3D morphology of these peritectic structures when viewing from different angles.

With an increase in the cooling rate to ~2.0 °C/s (Fig. 10b), a partly faceted morphology is observed (an intermediate regime). This is because the increased cooling rates disrupts the stability of the solute diffusion layer at the solid-liquid interface. In this regime, solid-liquid interface becomes unstable, and protrusions appear at the growing fronts, leading to a more frequent branching growth behaviour [17, 57, 61]. Concurrently, the kinetic effect starts to have influence because the time available for the peritectic reaction and coupled growth is reduced, resulting in smaller and chain-shaped peritectic $Al_6Mn$ phases with less perfect facets. The defects appear more fragmented and less regular, as the accelerated growth front outstrips the defect propagation speed.

Under the fast-cooling condition of ~20 °C/s (Fig. 10c), despite the very high thermodynamic driving force, the extremely short time prevent any sufficient solute diffusion to occur, hence the Mn-rich boundary necessary for sustaining the growth of $Al_6Mn$ shells cannot be established. Hence, the $Al_6Mn$ phase is forced to nucleate repeatedly and independently from the undercooled and solute-fluctuating melt [61]. This leads to much higher number of phases branching out with smaller and refined size (see Fig. 10c). The formed phases are much more segmented and rugged. The branches tangled together and most of them do not have sufficient time to follow a specific orientation relationship during growth. At such colling rate, the rapidly solidifying front disrupted any sustained solute and impurity accumulation that is necessary to initiate and maintain the formation of defects, hence these defects were largely suppressed. Therefore, cooling rate is one paramount processing parameter for controlling the peritectic phase size, orientation, morphology and defect formation.

## 5. Conclusion

Using quasi-simultaneous synchrotron X-ray diffraction and tomography, supplemented by EBSD and EDXS analysis, we have systematically studied *in-situ* and in real-time the nucleation and co-growth dynamics of the coupled peritectic phases in an Al-8wt%Mn alloy during the solidification process. The research provides a comprehensive 4D (3D space + time) dataset that help to reveal the underlying mechanisms of the peritectic reaction, in particular the critical role of the ~5 µm solute-rich boundary in the formation of the peritectic structures and the core defects. The results demonstrate that the final complex microstructure is not a random occurrence but is the deterministic outcome of a sequential process. The key findings of this research are:

(1) The primary $Al_4Mn$ phases (HCP structure) exhibited pronounced anisotropic growth kinetics. The growth rate was ~70 times higher along the <0001> direction than that along the <1$\bar{1}$00> direction, forming long hexagonal prisms at a cooling rate of 10 °C/min (~0.17 °C/s). A diffusion-controlled Mn-rich solute boundary layer with a thickness of ~5 µm was formed at the liquid-solid interface, serving as the precursor for growing the peritectic phase.

(2) At the peritectic reaction, the $Al_6Mn$ phase (orthorhombic structure) was epitaxially nucleated at the ~5 µm boundary layer in the slow cooling rate condition. The $Al_6Mn$ phases grew into thin shell surrounding the $Al_4Mn$ with the orientation relationship of $\{10\bar{1}0\}_{HCP}$ // $\{110\}_O$, $[0001]_{HCP}$ // $[001]_O$. Once the locally enriched Mn solute is depleted, the $Al_6Mn$ phase may nucleate and branch out at the edges and/or corners of the growing phase fronts, forming tetragonal prism-shaped structures that no longer follow the initial orientation relationship. The anisotropic solute diffusion also led to the formation of core defects at the growing centre of both $Al_4Mn$ and $Al_6Mn$ phases. Consequently, the final microstructures exhibit intertangled truss morphology in 3D.

(3) Cooling rate was the most critical parameter in driving the transition from a faceted to a non-faceted growth dynamic. With 0.17 °C/s, the peritectic structures had large and well-faceted morphologies with tubular defects along the growth axis. When the cooling rate was in the range of 2-20 °C/s, stable solute diffusion at the liquid-solid interface was disrupted, resulting in the formation of more refined, non-faceted, and rugged dendritic structures, with defects becoming further suppressed.

**Declaration of Competing Interest**

I declare that there is no conflict of interest among all authors concerning the submitted manuscript.

**Acknowledgements**

We would like to acknowledge the funding and financial supports given by the UK Engineering and Physical Science Research Council (EP/L019965/1), the National Natural Science Foundation of China (52104373) and Yunnan International Cooperation Base in Cloud Computation for Non-ferrous Metal Processing (202203AE140011). We also would like to acknowledge the allocation of synchrotron X-ray beam time (MG31637) on the DIAD beamline (K11) and electron beam time (MG35828) on the ePSIC beamline of the Diamond Light Source, UK, in particular Mr Matt Spink and Dr Mohsen Danaie, the free access to the University of Hull Supercomputer, Viper and the strong support from the technical team, in particular Mr Chris Collins. K. Xiang and Y. Wang would also like to acknowledge the financial support from the University of Hull and China Scholarship Council for their PhD studies at the University of Hull (202108500033 and 20220850023).

# Supplementary Materials

## 1. Al-Mn binary phase diagram and the peritectic reaction of Al-8wt.%Mn alloy

Fig S1 shows the Al-Mn equilibrium phase diagram [1]. The peritectic reaction of the Al-8wt%Mn alloy occurs at 705 ºC, and the three phases are (Mn in atomic percentage):

**Liquid (~2.5 at.% Mn) + µ-Al$_4$Mn (20 at.% Mn) = Al$_6$Mn (14.3 at.% Mn)**

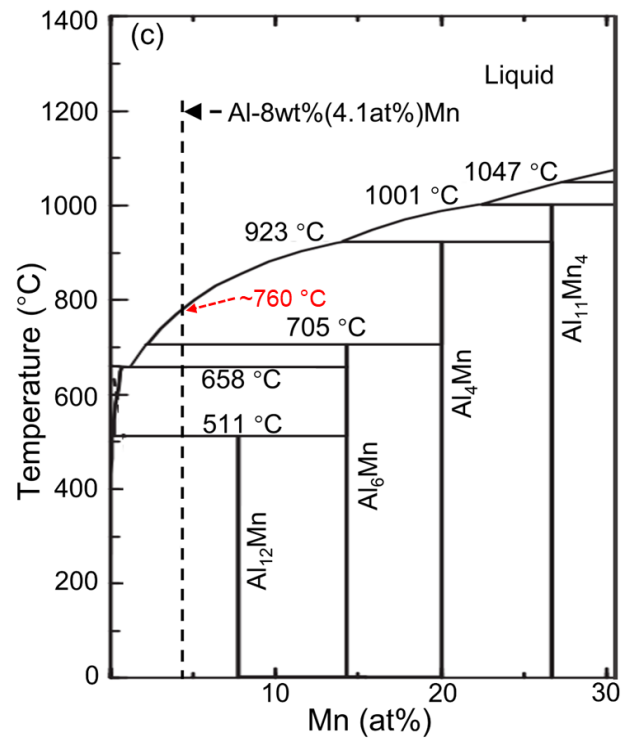

Fig. S1. Al-Mn phase diagram (to 30 at.%Mn) [1]. The dash line marks the Al-8wt.%Mn alloy used.

## 2. Ex-situ electron microscopy characterization of the peritectic microstructures

After the in-situ synchrotron X-ray experiments, the microstructures and elemental distribution of the solidified Al-8wt% Mn alloy samples were ex-situ characterized using electron microscopy. Firstly, a ~3 mm long rod bar was cut from the middle section of the solidified sample and then hot-press mounted using conductive Bakelite powders (MetPrep Ltd.) inside a hot mounting machine (BUEHLER UK Ltd.). The mounted sample was ground and polished according to the Five-Step Contemporary Procedure [2]. The microstructures were characterized using a JSM-6610 Scanning Electron Microscope (SEM, JEOL Ltd.) equipped with an Energy Dispersive X-ray Spectroscopy (EDXS) detector (Oxford Instruments Xmax$^n$ 80T) and an Electron Backscattered Diffraction (EBSD) detector (Oxford Instruments NordlysNano) available at the Research Complex at Harwell. The EBSD scans were conducted at 20 kV with a 1 µm step size. The collected EBSD data were analysed using the Aztec 2.4 software and HKL CHANNEL5 (v5.0.9.0) software. Fig. S2a (a backscatter electron image) shows the microstructures of the solidified sample in a typical 2D area, and Fig. S2b shows the zoom-in image for the peritectic phases inside the dotted-line square in Fig. S2a, consisting of the primary $Al_4Mn$ phase in the central region and the reacted peritectic phase $Al_6Mn$ surrounding the $Al_4Mn$.

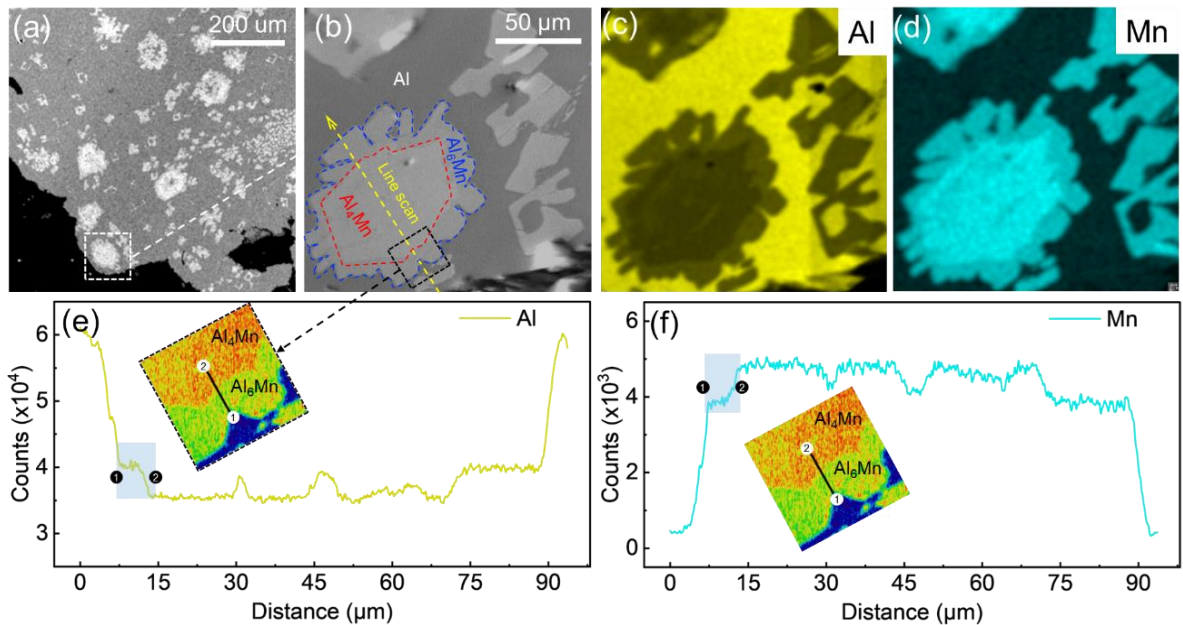

Fig. S2. (a) A typical SEM-BSE image of the solidified Al-8wt.%Mn sample after the in-situ experiment. (b) The enlarged image for the area within the dotted-line square in (a), showing the typical coupled peritectic structures. (c) and (d) EDXS elemental map of Al and Mn respectively. (e) and (f) The EDXS line scan of Al and Mn along the yellow dashed line in (b). The insets in (e) and (f) show the enlarged view of the region marked by the white dotted-line square in (b).

Figs. S2c and d show the Al and Mn elemental maps, respectively. Figs. S2e (for Al) and f (for Mn) are the line scans along the yellow dotted line in Fig. S2b. showing the count of Al and Mn across the coupled peritectic structure within the white dotted-line square shown in Fig. S2b, which was further converted into a composition pseudo colour map, enlarged and then inserted in Fig. S2e for easy visualisation. The count profiles for the line segment between point 1 and point 2 in Figs. S2e and f indicate the change of Al and Mn respectively across the $Al_6Mn$-$Al_4Mn$ interface (a measured thickness of ~5 μm as highlighted by the shaded areas in Figs. S2e and f, respectively).

Table S1. The EDXS-measured and equilibrium compositions (shown in the phase diagram) of the Al4Mn and Al6Mn phases.

| Phase | EDXS measured | | Equilibrium composition in the phase diagram | |
|---|---|---|---|---|
| | Al (at.%) | Mn (at.%) | Al (at.%) | Mn (at.%) |
| Al4Mn | 81.42 | 18.58 | 80 | 20 |
| Al6Mn | 85.88 | 14.12 | 85.7 | 14.3 |

Multiple EDXS point quantitative analyses (see Table S1) were made at the homogeneous areas for the $Al_4Mn$ and $Al_6Mn$ phases and determined that the Mn is ~18.58 at.% for the $Al_4Mn$ and ~14.12 at.% for the $Al_6Mn$, which agreed very well with the equilibrium compositions shown in the phase diagram. However, the EDXS line-scan across the $Al_6Mn$-$Al_4Mn$ interface indicated that there existed a gradual change of Mn concentration (or Al, see the shaded areas between point 1 and point 2 in Figs. S2e and f) across the interface. Such gradual composition change across the interface was also reflected in the grey value change in the 2D projections of the tomography data, as described in detail below.

3. **Gradual change of the pixel grey values across the $Al_6Mn$-$Al_4Mn$ interface in 2D projection of the tomographic data.**

Fig. S3a1 shows the tomography of a typical long hexagonal $Al_4Mn$ primary phase (the precursor prior to the peritectic reaction) at 706 °C. After completion of the peritectic reaction at 688 °C, the precursor was wrapped around by a thin layer of $Al_6Mn$ phase (see Fig. S3b1). Figs. S3a2 and b2 show the 2D slices taken from shaded planes at the positions shown in Fig. S3a1 and b1. Along the dotted lines in Figs. S3a3 and b3, the pixel grey values are plotted and shown in Fig. S3c. Clearly, for the $Al_4Mn$ (whether in the $Al_4Mn$ precursor or in the $Al_4Mn$ after the peritectic reaction), the averaged grey values are relatively constant at ~200 a.u. However, across over the $Al_6Mn$-$Al_4Mn$ interface along the black line in Fig. S3b3 (both ends of the black line are highlighted with black dots), there are clear changes of the grey value with the profile's slope in the blue-shaded region matching the profile's slope of the EDXS line-scan (see the profiles in Figs. S2e and f).

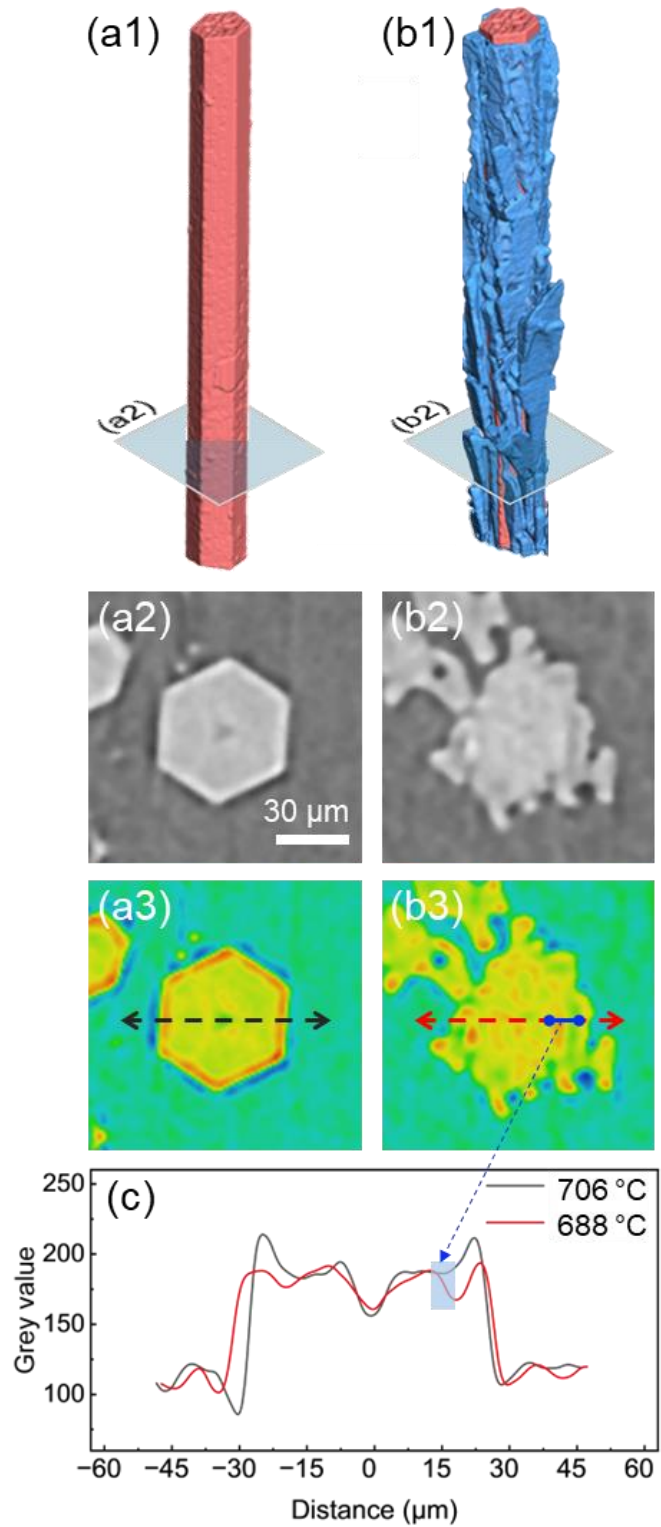

Fig. S3. (a1) tomography of the peritectic precursor $Al_4Mn$, and (b1) the peritectic product $Al_6Mn$. (a2 and a3) 2D slices and the corresponding pseudo-colour attenuation maps of the $Al_4Mn$ precursor at the positions indicated by the shaded squares in (a1). (b2 and b3) 2D slices and the corresponding pseudo-colour attenuation maps of the peritectic $Al_6Mn$ at the positions indicated by the shaded squares in (b1). (c) Grey-value profiles taken along the dotted lines in (a3) and (b3).

## 4. Establishing the correlation between the Mn concentration and the pixel grey values

Fig. S4a superimposes the EDXS profile for the Mn and the grey-value profile across the Al$_6$Mn-Al$_4$Mn interface (the interface was enlarged for a clearer view).

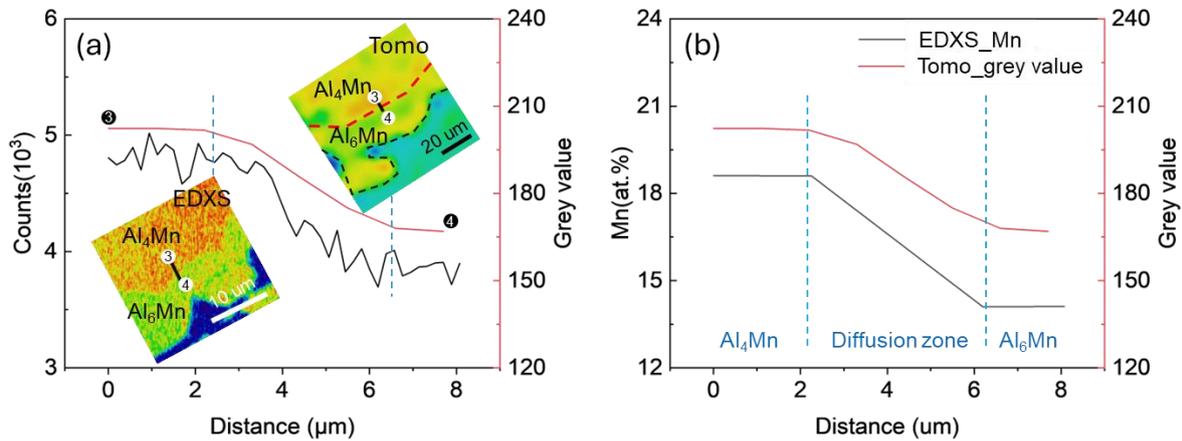

Fig. S4. (a) Comparison of the EDXS Mn-count profile with the grey-value profile extracted from the tomographic slice across the Al$_6$Mn-Al$_4$Mn interface (the line segment between point 3 and point 4). The left inset is the one in Fig. S2e. The right inset is the white dotted square in Fig. S3b3). (b) The converted Mn concentration profile (the black profile after a data smooth operation) with the grey-value profile (red) across the interface (the line segment between point 3 and point 4).

Clearly, across the interface (see the inset from point 3 to point 4), there exists a Mn diffusion layer of ~5 μm thick with nearly the same slope for the EDXS line scan count and the grey value profile from the tomography slice. Such a good agreement confirmed that the change of the grey value across the interface actually reflects the change of Mn concentration. Hence, we converted (via a normalisation operation) the EDXS Mn count to the atomic percentage of the Mn in the Al$_4$Mn and Al$_6$Mn (see Fig. S4b). The Mn (18.58 at.%) in the Al$_4$Mn corresponds to a grey value of 200 a.u (2D slice), whereas 14.12 at.% Mn in the Al$_6$Mn phase corresponds to a grey value of 167 a.u. Then a linear relationship between the Mn concentration $C_{Mn}$ (at.%) and grey value G was established across the Al$_6$Mn-Al$_4$Mn as below:

$$C_{Mn} = 0.133\,G - 8.296 \tag{S1}$$

Since all in-situ tomograms in this work were acquired from the same Al-8wt.%Mn alloy under identical imaging conditions and the precursor $Al_4Mn$ phases have near-constant grey value during solidification, Eq. S1 was applied to all datasets to obtain the Mn concentration profiles across the $Al_4Mn$ phase and the surrounding liquid.

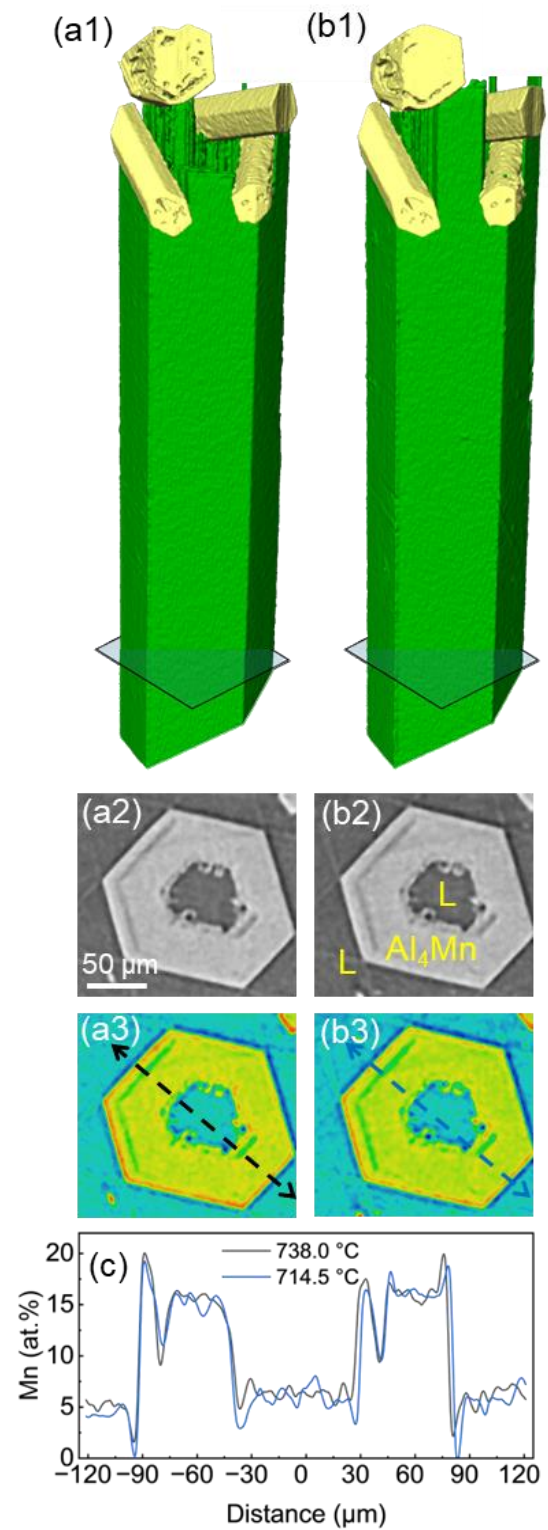

Fig. S5. (a1 and b1) 3D morphological evolution of a typical $Al_4Mn$ phase during solidification. (a2 and b2) The typical 2D slices and (a3 and b3) the corresponding pseudo-colour attenuation maps of $Al_4Mn$ phase at the cross-sections indicated by the shaded squares in (a1 and b1), respectively. (c) Calibrated Mn-concentration profiles along the dashed lines in (a3 and b3), calculated using Eq. S1.

Figs. S5a1 and b1 show the 3D morphological evolution of a typical $Al_4Mn$ primary phase during solidification from 724 °C to 706 °C (prior to the peritectic reaction, see Fig. S1). The equilibrium phase diagram (Fig. S1) indicates that, over this temperature range, the Mn in the liquid decreased from ~4 at.% to 3 at.%, while that in the $Al_4Mn$ phase decreased from 20 at.% to 19 at.%. Using Eq. S1, the calculated Mn concentration in the liquid (Fig. S5c) decreases from ~6 at.% to 4 at.%, whereas the Mn content in the $Al_4Mn$ phase is in the range of 20 at.% to 16 at.%. The broadly consistent trend and magnitude between the calculated values and the phase-diagram predictions indicate that the approach we used is sensible and consistent.

## 5. EBSD analyses of the orientation relationship of the peritectic structure

To determine the orientation relationship of the $Al_4Mn$ and $Al_6Mn$ in the peritectic structure, the coordinate system of the samples was adjusted in the virtual chamber to allow the basal plane (i.e., {0001}) of the $Al_4Mn$ phase to be horizontally perpendicular to the screen. It should be noted that this does not change the relative orientation relationship between the $Al_4Mn$ and $Al_6Mn$ phases. By comparison, the highly overlapping poles undoubtedly demonstrate the existence of a specific crystallographic orientation relationship between the peritectic precursor ($Al_4Mn$ of HCP structure) and its products ($Al_6Mn$ of orthorhombic structure), i.e., $\{10\text{-}10\}_{HCP}$ // $\{110\}_O$, $[0001]_{HCP}$ // $[001]_O$.

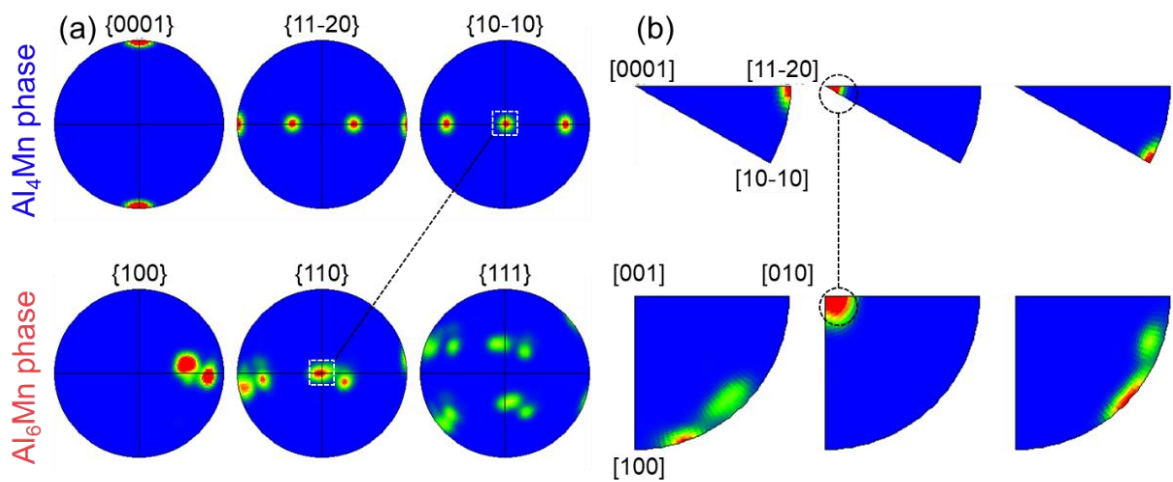

Fig. S6. (a) The pole figures and (b) the inverse pole figures of the $Al_4Mn$ and $Al_6Mn$ phases.